\documentclass[aps,pra,twocolumn,showpacs,preprintnumbers,amsmath,footinbib,10pt,longbibliography]{revtex4-2}

\pdfoutput=1
\makeatletter
\renewcommand*{\@fnsymbol}[1]{\ensuremath{\ifcase#1\or \or \spin\or \covbond\or \boson\or \magnon\or \atom\or \*\or \quadrupole\or
   \mathsection\or \mathparagraph\or \|\or **\or \dagger\dagger
   \or \ddagger\ddagger \else\@ctrerr\fi}}
\makeatother

\usepackage{hyperref}
\usepackage[utf8]{inputenc}
\usepackage[sort&compress]{natbib}
\usepackage[dvipsnames]{xcolor}
\usepackage{graphicx}
\usepackage{epsfig}
\usepackage{epstopdf}
\usepackage{amsfonts}
\usepackage{amstext,amsmath,amssymb}
\usepackage{breqn}
\usepackage{soul}
\usepackage{rotating}
\usepackage{dcolumn}
\usepackage{array,multirow}
\usepackage[english]{babel}
\usepackage[export]{adjustbox}
\usepackage{braket}
\usepackage[caption=false,subrefformat=parens,labelformat=parens]{subfig}
\usepackage{xr}
\usepackage{ifpdf}
\usepackage{tikz}
\usepackage{svrsymbols}
\usetikzlibrary{arrows}

\hypersetup{urlcolor=violet, citecolor=violet, linkcolor=violet, colorlinks=true}
\newcommand{\be}{\begin{equation}}
\newcommand{\ee}{\end{equation}}
\newcommand{\bea}{\begin{eqnarray}}
\newcommand{\eea}{\end{eqnarray}}

\usepackage{color}

\newcommand{\santi}[1]{{\color{black} {\it} #1}}

\makeatletter
\let\cat@comma@active\@empty
\makeatother

\begin{document}

\title{Low frequency signal detection via correlated Ramsey measurements}

\author{Santiago Oviedo-Casado${}^{\magnon,\atom,\quadrupole}$, Javier Prior${}^{\covbond,\spin}$, Javier Cerrillo${}^{\magnon,\boson}$}

\affiliation{${}^{\magnon}$\'Area de F\'isica Aplicada, Universidad Polit\'ecnica de Cartagena, Cartagena E-30202, Spain\\
${}^{\atom}$Racah Institute of Physics, The Hebrew University of Jerusalem, Jerusalem 91904, Givat Ram, Israel\\
${}^{\covbond}$Departamento de Física - CIOyN, Universidad de Murcia, Murcia E-30071, Spain
}

\email{${}^{\quadrupole}$oviedo.cs@mail.huji.ac.il\\
${}^{\spin}$javier.prior@um.es\\
${}^{\boson}$javier.cerrillo@upct.es}

\begin{abstract}
The low frequency region of the spectrum is a challenging regime for quantum probes. We support the idea that, in this regime, performing Ramsey measurements carefully controlling the time at which each measurement is initiated is an excellent signal detection strategy. We use the Fisher information to demonstrate a high quality performance in the low frequency regime, compared to more elaborated measurement sequences, and to optimise the correlated Ramsey sequence according to any given experimental parameters, showing that correlated Ramsey rivals with state-of-the-art protocols, and can even outperform commonly employed sequences such as dynamical decoupling in the detection of low frequency signals. Contrary to typical quantum detection protocols for oscillating signals, which require adjusting the time separation between pulses to match the half period of the target signal, and consequently see their scope limited to signals whose period is shorter than the characteristic decoherence time of the probe, or to those protocols whose target is primarily static signals, the time-tagged correlated Ramsey sequence simultaneously tracks the amplitude and the phase information of the target signal, regardless of its frequency, which crucially permits correlating measurements in post-processing, leading to efficient spectral reconstruction.
\end{abstract}


\maketitle

\section{Introduction}
\label{sec:introduction}
 Identifying weak signals in the low frequency regime of the electromagnetic spectrum is paramount for the study of a wide variety of physical systems, ranging from low energy fundamental particles, such as neutrinos or axions \cite{CERN}, to chemical bonds (J-couplings) \cite{Soncini2003,Sutter2012}, and material defects \cite{Bevington2019}. As a consequence, the last years have witnessed an effort to develop quantum sensors operating in this regime, which could be revolutionary for quantum information processing \cite{Majety2022}, quantum communications and radar applications \cite{Gerginov2017,Quntao2022}, biomedical imaging \cite{Xia2006,Belfi2007,Kim2014} or nuclear quadrupole resonance \cite{Garroway2001,Lee2006}. However, existing sensors based on, e.g., Rydberg atoms \cite{Gordon2014,Meyer2018}, masers \cite{Jiang2021}, optomechanical sensors \cite{Bagci2014,Rudd2019}, or superconducting circuits \cite{Gely2019}, are restricted by sophisticated setups or extreme operating temperatures. 

The typical quantum sensing setup features a carefully controlled quantum probe prepared on an initial coherent superposition state. Interaction with an external signal alters the quantum trajectory of the probe, making its final state dependent on a number of parameters --such as the frequency $\omega$, the phase $\varphi$, or the amplitude $\xi$-- characteristic of the external signal, which is the underlying principle of quantum sensing. Repeating the measurement process to gather statistics allows to estimate the signal parameters. Yet this very same measurement process means that the system interacts with a noisy environment that hinders the performance of the quantum sensor. Such noise causes two distinct processes, namely, dephasing or $T_2^*$ noise, that describes the characteristic survival time of a coherent superposition state, and relaxation or $T_1$ noise, which is the time the qubit takes to return to its equilibrium state \cite{Barry2020}. Noise imposes severe constraints on the ability to detect signals whose frequency is smaller than the inverse characteristic dephasing time \cite{Joas2017}. Here, we challenge this limitation.

During the evolution time, control sequences are applied to ensure sensitivity to the signal of interest, and to avoid interaction with unwanted noise. The Ramsey sequence lets the probe evolve freely, making it most sensitive to the slowest frequency component of the noise spectrum, for which reason Ramsey spectroscopy is traditionally used to estimate parameters from static signals. In a Ramsey measurement, the probe is sensitive to all noise sources and, consequently, it is only useful if its interrogation time $\tau_R$ is such that $\tau_R < T_2^*$. Detecting oscillating signals requires suppressing the sensitivity to the noisy frequencies lower than that of the target signal, $\omega$, which is most commonly achieved through a family of control sequences collectively known as dynamical decoupling (DD) \cite{Viola1999,Cywinski2007,Degen2007}. They comprise pulses intended to refocus the probe and to filter out noise frequencies below $\omega_{DD} = \pi/\tau$, with $\tau$ the pulses separation \cite{Cywinski2007}. This has two consequences: on the one hand, DD demands $\omega_{DD} \approx \omega$, requiring previous knowledge of the target signal frequency, and fixing upfront the duration of the corresponding sequence.  On the other hand, DD extends the decoherence time of the qubit to $T_2 \geq T_2^*$ \cite{Pham2012a,Barry2020}, which is typically determined experimentally \cite{deLange2010}. Both facts, when put together, mean that the spacing between pulses in DD cannot exceed $T_2$, posing a problem for low frequency signals $\omega < \pi/T_2$, as DD sequences cannot eliminate the noise around the target signal's frequency. Moreover, although there are techniques that can be embedded within DD sequences, and that permit extending the $T_2$ of the probe \cite{Cai2012}, these sequences are ultimately limited by $T_2 < T_1$. The immediate consequence being that sensing sufficiently low frequencies is not possible with DD.

In this article, we explore the correlated Ramsey protocol for quantum parameter estimation of low frequency signals, which uses the quantum heterodyne (Qdyne) technique \cite{Schmitt2017,Degen2017,Glenn2018} to make the Ramsey sequence suitable for oscillating signals, and that was initially proposed by Herbschleb et al. \cite{TheOther}. There, the authors demonstrate experimentally the performance of the correlated Ramsey protocol for low frequency detection using bulk NV-centres, which feature $T_2^*$ times in the order of milliseconds. Rather, we use a distinct objective tool, namely, the Fisher information, to study the performance of the correlated Ramsey sequence, comparing it to its immediate mathematical kindred, i.e. dynamical decoupling, which permits us revealing the reason for the success of correlated Ramsey in the low frequency region of the spectrum. Further, the Fisher information allows us to optimise the correlated Ramsey sequence according to the particular features of any given experiment, thus showing unambiguously the excellent performance that it has in the typical regime of nuclear magnetic resonance at the nano-scale, which requires having probes capable of being very close to the sample, with the consequence that they feature $T_2^*$ times in the much more restrictive microsecond regime, which then requires a different optimal sequence design, as we demonstrate. The use of the Fisher information permits us to characterise and optimise the parameters defining the sequence, such that it can be tailored to the particular experiment and parameter of interest; for example, the frequency. Additionally, we compare the performance of correlated Ramsey to state-of-the-art low frequency protocols such as continuous wave and pulsed ODMR \cite{Acosta2010,Schoenfeld2011,Dreau2011,Clevenson2015,Schloss2018}, using the frequency detection capabilities as a benchmark.

\begin{figure}
\includegraphics[width=\columnwidth]{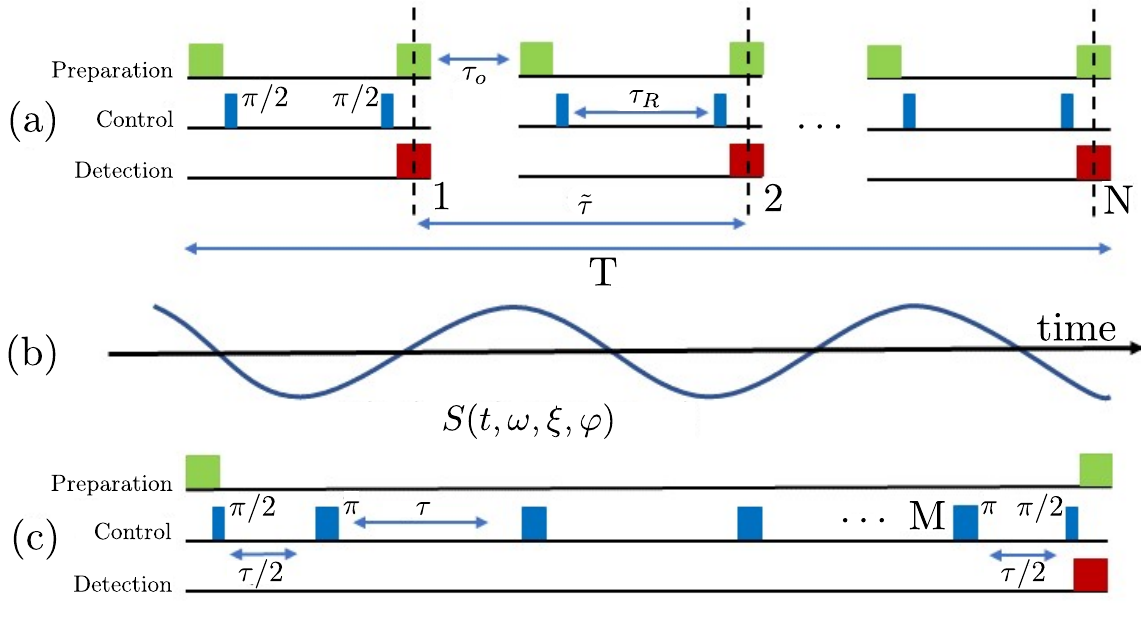}
    \caption{(a) Correlated Ramsey protocol composed of $N$ Ramsey measurements repeated sequentially every $\tilde\tau$, which includes the measurement time $\tau_R$ and an overhead time $\tau_o$ for qubit readout and initialization, such that the total duration of the protocol is $T = N\tilde\tau$. (b) Low frequency target signal $S(t,\omega,\xi,\varphi)$. (c) Dynamical decoupling sequence of equivalent duration $T = M\tau$ featuring $M$ $\pi$ pulses with even time separation $\tau$. For a low frequency, a series of equally spaced Ramsey measurements imitate the behaviour of a dynamical decoupling sequence of equivalent duration, avoiding, at the same time, having to match the half period of the signal to the pulses separation, thereby vanquishing the problem of the probe's decoherence time while keeping the benefits of DD sequences}.\label{Fig1}
\end{figure}

Typically, detection of oscillating signals requires a frequency matching  $\omega_{DD} \approx \omega$, such as in dynamical decoupling, or fluorescence variations detection, as in ODMR. The former is limited by the coherence time of the probe, while the later has a sub-optimal scaling with the total measurement time. Correlated Ramsey takes advantage of the best of both worlds to simultaneously track the amplitude and the phase of the oscillating signal, permitting correlating the measurements during post-processing, thereby having a better scaling with the total measurement time unburdened by any frequency matching condition, thus avoiding the problem of the limited coherence time of the probe. The correlated Ramsey protocol becomes sensitive to the {\textit{coherent}} phase of the target signal, making it distinguishable from  {\textit{incoherent}} noisy background signals of similar frequency. This fact has already been taken advantage of to propose Qdyne-like sequences capable of detecting highly oscillating fields \cite{Cai2021}, to which similar Ramsey-like synchronized measurements have also  been applied \cite{Meinel2021,Staudemaier2021}. In this article, we show explicitly that it is through using the information about the target signal's phase, that the Qdyne measurement protocol acquires, what allows a Ramsey sequence to be maximally sensitive to oscillating signals, behaving as a dynamical decoupling sequence whose specific target is the low frequency regime, overcoming the limitations to DD sequences imposed by the coherence time of the probe in that regime. We show theoretically the excellent performance of the correlated Ramsey protocol in the low frequency regime, and propose the best experimental parameters to design optimal sequences. 

\section{Theory}
\label{sec:theory}

\subsection{Formal definitions}

Consider a qubit probe interacting with the longitudinal component of an external signal $S(t,\omega,\xi,\varphi)$ [see Fig.~\ref{Fig1}(b)]. The interaction may be modeled by the Hamiltonian $\sigma_z S(t,\omega,\xi,\varphi)$, with $\sigma_z = \ket{1}\bra{1}-\ket{0}\bra{0}$. The probe is additionally subject to transversal noise, which we treat through the Lindblad master equation formalism. Then, an initial superposition state on the probe experiences decoherence with a characteristic time $T_s$ that depends on the environment of the qubit and on the control sequence, $s$, used.  For Ramsey, $T_{R} = T_2^*$, and $T_{DD} = T_2$ for DD. The probability to find the qubit in its original state after some evolution time $t$ is
\be
P_s = \frac{1+e^{-t/T_s}\cos\left[\Phi_{s}\left(t,\omega,\xi,\varphi\right)\right]}{2},
\label{eq:P}
\ee
where $\Phi_{s}(t,\omega,\xi,\varphi)$ is the accumulated phase that depends on the specific sequence, the evolution time $t$, and the frequency $\omega$, the amplitude $\xi$, and the phase $\varphi$ of the external signal.

We want to find the optimal control sequence that yields the smallest mean squared error $\Delta \omega$ in the estimation of any signal parameter, e.g. the frequency $\omega$, for a fixed experiment duration. According to the Cram{\'e}r-Rao bound, the Fisher information ($I_s^\omega$) contained on a sequence $s$ about the frequency bounds $\Delta \omega$, such that $\Delta \omega \geq 1/I_s^\omega$ \cite{Wootters1981,Braunstein1994}. We can then compare the performance of different sequences by computing their Fisher information with \cite{Gefen2017}
\begin{equation}
I_s^\alpha = \frac{1}{P_s(1-P_s)}\left(\frac{dP_s}{d\alpha}\right)^2,
\end{equation}
with $\alpha$ referring to any of the parameters to be estimated. Considering Eq.~(\ref{eq:P}), $I_s^\alpha$ relates to $\Phi_s$ through:
\be
I_s^\alpha = \frac{\sin^2\left[ \Phi_s(t,\omega,\xi,\varphi)  \right]}{\exp\left(2 t/T_s\right) - \cos^2\left[ \Phi_s(t,\omega,\xi,\varphi) \right]}\left[ \frac{d\Phi_s(t,\omega,\xi,\varphi)}{d\alpha} \right]^2.
\label{FIone}
\ee
Repeated experiments additively increase the Fisher information, such that for $N$ measurements spanning a total experiment time $T$ 
\begin{equation}
    I_{Ns}^\alpha(T,\omega,\xi,\varphi)=\sum_{j=1}^N I_s^\alpha(\tau_s,\omega_j,\xi_j,\varphi_j),
    \label{eq:FIrepeated}
\end{equation}
where we allow for modification of the external parameters for different experimental realizations.

To gain insight on the performance of correlated Ramsey, we compare  the performance of a single DD experiment featuring $M$ $\pi$ pulses, and of total duration $T = M\tau$, with $N$ Ramsey experiments of total equivalent duration $T = N\tilde\tau$,
where $\tilde\tau$ is the sum of the duration of the experiment $\tau_R$ and an overhead time $\tau_o$, as depicted in Fig.~\ref{Fig1}. To do so, we resort to the gain  
\begin{equation} 
    g^\alpha\equiv\frac{I_{NR}^\alpha(T)}{I_{DD}^\alpha(T)}.\label{eq:exactgain}
\end{equation} 
In the limit of low frequencies we show that, for a wide range of parameters, $g^\alpha$ grows exponentially as the frequency of the external signal decreases, demonstrating that correlated Ramsey experiments are a suitable experimental procedure for low frequency sensing.

\subsection{Single Ramsey experiment}

We begin by considering the Ramsey sequence, which consists on the free evolution of the qubit over a time $\tau_R$. The qubit is initially prepared on a superposition state $\Psi(0) = (\ket{0} + \ket{1})/\sqrt{2}$ by means of a $\pi/2$ pulse. A final $\pi/2$ pulse rotates the qubit back to the measurement basis. Note that all pulses will be considered here as having negligible duration (also known as the impulsive limit) and with amplitude much smaller than the energy gap of the qubit. During the free evolution time, the superposition accumulates a phase $\Phi_R(t,\omega,\xi,\varphi) = \int_{0}^{t}dt' S(t',\omega,\xi,\varphi)$
due to the external signal. Assuming that $S(t,\omega,\xi,\varphi) = \xi\cos(\omega t + \varphi)$, at the end of the sequence the total accumulated phase is
\be
\Phi_R(\tau_R,\omega,\xi,\varphi) = \xi\tau_R\cos\left(\frac{\omega\tau_R}{2} + \varphi\right){\text{sinc}}\left(\frac{\omega\tau_R}{2}\right).
\label{acphaser1}
\ee
Typically, sensitivity to external signals is quantified through the filter function (FF) of a given sequence, defined as $\langle \Phi_R(\tau_R,\omega,\xi,\varphi)^2\rangle_{\varphi}$, the signal's phase averaged square of the accumulated phase, and calculated in terms of its parameters alone \cite{Biercuk2011}. Nevertheless, in the context of our discussion, it can also be interpreted as $\langle \Phi_R(\tau_R,\omega,\xi,\varphi)^2\rangle_{\varphi}$, the brackets denoting an average over the signal's phase. For Ramsey spectroscopy, the FF reads $F_R(\omega) = \xi^2\tau_R^2{\text{sinc}}^2\left(\omega\tau_R/2\right)/2$ \cite{Cerrillo2021}, which peaks at vanishing frequency, supporting the idea that Ramsey sequences are mostly sensitive to static signals. Yet Eq.~(\ref{acphaser1}) indicates that this need not be the case if the phase of the signal is taken into account. For measurements performed at random initial times, information about $\varphi$ is lost, and the relevant equation describing an experimental outcome is the FF of the given sequence. If the starting time $t_j$ of each sequence is recorded, we can write $\varphi_j = \omega t_j +\varphi$, such that information about $\varphi$ is conserved. A suitable measurement can then retrieve this information through the dependency in the phase expressed in Eq.~(\ref{acphaser1}) (see the Appendix \ref{Methodsphase} for details), revealing the deep connection existing between filter functions and the phase that a given probe accumulates due to the interaction with an external signal. A relation that can be used to tailor control sequences for given signals. 

\subsection{Dynamical decoupling}

DD sequences are constructed by embedding $\pi$ pulses in between initialization and readout of the qubit. These pulses invert the state of the qubit with respect to a specific axis on the Bloch sphere. The effect of $M$ $\pi$ pulses on the total accumulated phase can be computed by means of a response function $h_M(t)$, which has a unit absolute value, but changes sign every time a $\pi$ pulse is applied.
Then, the accumulated phase on the qubit is the convolution of $h_M(t)$ and the external signal $S(t,\omega,\xi,\varphi)$ \cite{Cerrillo2021}, yielding
\begin{equation}
\begin{split}
\Phi_{DD}(T,\omega,\xi,\varphi) &= -\xi\tau\cos\left(M\frac{\omega\tau + \pi}{2} + \varphi \right) \\&
\frac{\sin\left(M\frac{\omega\tau+\pi}{2}\right)}{\cos\left(\frac{\omega\tau}{2}\right)}\sin\left(\frac{\omega\tau}{4}\right){\text{sinc}}\left(\frac{\omega\tau}{4}\right).
\label{acphasehn}
\end{split}
\end{equation}
DD sequences create a filter around their characteristic frequency $\omega_{DD} = \pi/\tau$, whose width is inversely proportional to the number of pulses. 
Such a filter effect can be better understood studying the accumulated phase around the maximally sensitive region. Expanding Eq.~(\ref{acphasehn}) around $\delta = \omega -  \omega_{DD} \ll \omega$ the accumulated phase yields $\Phi_{DD}(T,\delta,\xi,\varphi) \sim 2\Phi_R(T,\omega,\xi,\varphi)/\pi$, showing that DD is equivalent to a Ramsey sequence if the frequency variable is expressed in the adequate frame of reference i.e. a DD sequence is most sensitive to the slowest frequency component of the noise spectrum as seen by the rotating qubit. Conversely, a correlated Ramsey sequence is, effectively, a dynamical decoupling protocol not limited by the frequency matching condition.


\section{Results}
\label{sec:results}

\subsection{Fisher information}

 All equations above are exact within the impulsive limit. In order to demonstrate the sequence optimisation procedure, and to easily interpret the (also exact) numerical results ahead, we now derive an approximate expression for the low frequency limit of the accumulated phases and the gain $g^\alpha$. In the low frequency limit, most terms resulting from the derivative in Eq.~(\ref{FIone}) vanish, and we keep only those proportional to the ${\text{sinc}}$ function. In DD, the first term in Eq.~(\ref{FIone}) can be upper bound by $\exp{(-2T/T_2)}$. Similarly, correlated Ramsey is upper bound by $\exp{(-2\tau_R/T_2^*)}$. Note that, in correlated Ramsey, it is $\tau_R$, the duration of a single measurement, and not $T$, what upper bounds the Fisher information. The reason is that in correlated Ramsey, dephasing affects each measurement individually and not the global sequence, as the qubit is repolarised into its $\ket{0}$ state after each measurement, which means that the exponential is a common upper bound of all terms in Eq.~(\ref{eq:FIrepeated}). Then, the sum in Eq.~(\ref{eq:FIrepeated}) can be simplified noting that the derivatives of oscillating functions can be written as phase-shifts in Eq.~(\ref{acphaser1}).

Considering these approximations for the low frequency regime, and keeping only the leading order in the total measurement time $T$ (see Ref.~\cite{Schmitt2017} and Appendix \ref{MethodsFI} for details of the calculations), for correlated Ramsey
\begin{dmath}
I_{NR}^\alpha(T) = f_{NR}^\alpha(T)\frac{\tau_R^2 T}{2\tilde\tau}
e^{-2 \tau_R/T_2^*},
\label{FIRamsey}
\end{dmath}
with $f_{s}^\alpha$ a factor that depends on the specific target parameter and sequence $s$ employed. Here $f_{NR}^\omega = \xi^2 T^2/3$, $f_{NR}^\xi = 1$, and $f_{NR}^\varphi = \xi^2$. alternatively, for a single DD measurement
\begin{equation}
I_{DD}^\alpha(T) = f_{DD}^\alpha(T)\frac{T^2}{2\pi^2} 
e^{-2T/T_2},
\label{FIDD}
\end{equation}
with $f_{DD}^\omega = 3f_{NR}^\omega$ and $f_{DD}^\xi = 4f_{NR}^\xi$. Note that, to fairly compare with correlated Ramsey, we perform a phase averaging in Eq.~(\ref{FIDD}) which results in $I_{DD}^\varphi$ = 0 and that decreases the total Fisher information by a factor of two. Dividing Eq.~(\ref{FIRamsey}) by Eq.~(\ref{FIDD}) 
gives us
\begin{equation}
    \lim_{\omega\rightarrow0}g^\alpha=e^{2\left(\frac{T}{T_2}-\frac{\tau_R}{T_2^*}\right)}\frac{\pi^2\tau_R^2}{T\tilde\tau}f^\alpha,\label{lowfreqgain}
\end{equation}
with $f^\omega=1/3$ and $f^\xi = 1/4$.

\begin{figure}\includegraphics[width=\columnwidth]{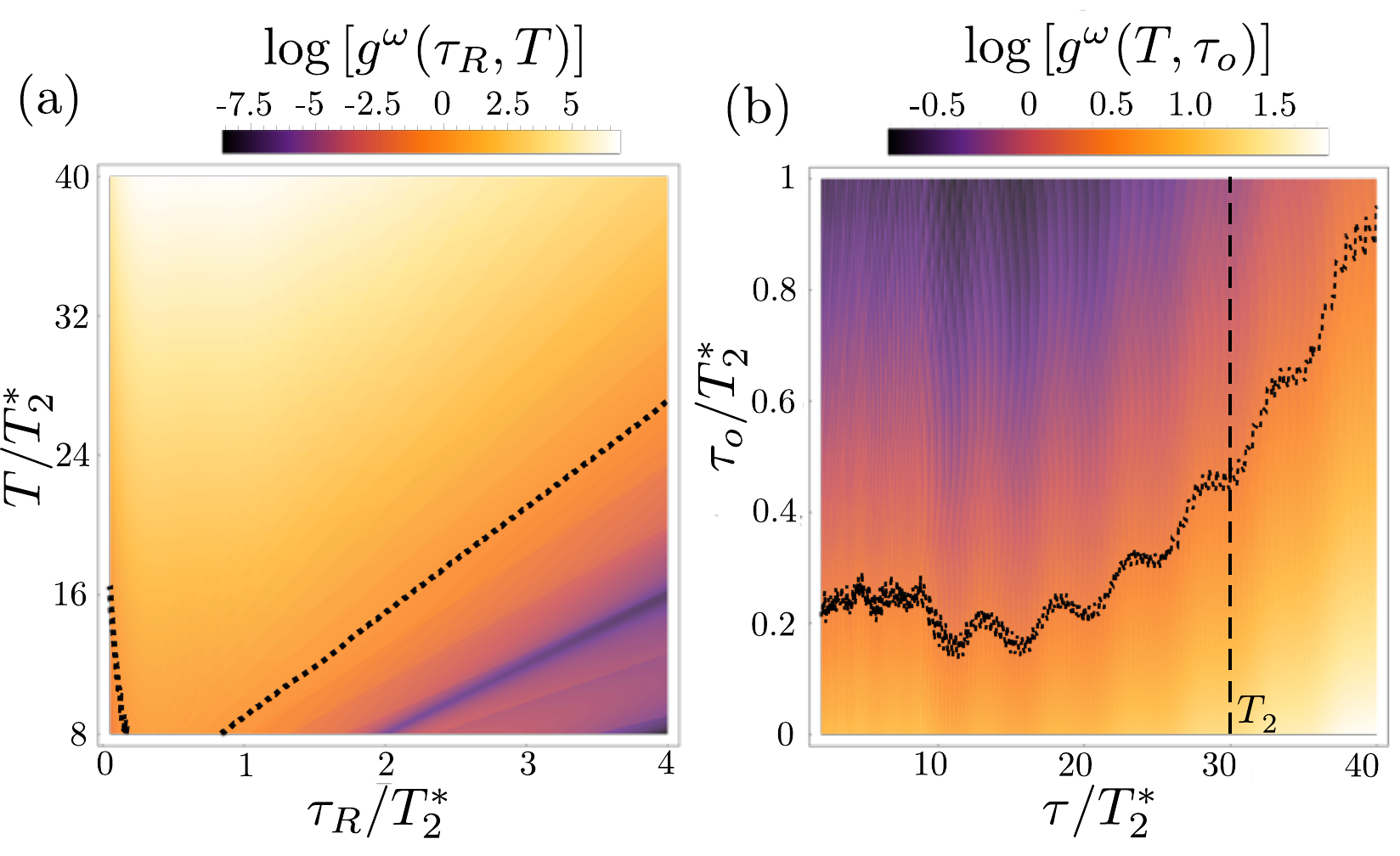}
   \caption{Logarithm of the exact gain for frequency estimation $g^\omega$ from Eq.~(\ref{eq:exactgain}). In (a), we compare correlated Ramsey to a DD sequence of eight pulses, which we assume to provide a coherence time enhancement $T_2 = 8T_2^*$. We set the duration of a single Ramsey measurement to $\tilde\tau = 0.5T_2^* + \tau_R$, which includes both the phase acquisition time $\tau_R$ and the overhead time $\tau_o$. We plot $g^\omega$, the gain corresponding to the frequency, as a function of $T$, the duration of either experiment, and of $\tau_R$. In (b), we compare correlated Ramsey to a \santi{DD} sequence for which we assume a hypothetical strong recoherence effect of $T_2 = 30T_2^*$, \santi{marked by the vertical dashed line}. The total length of the sequence,  \santi{$T = M\tau$, is optimized through the number of pulses $M$ to yield the highest phase accumulation possible for a given $\tau$ that}  ranges from $2T_2^*$ to $40T_2^*$. In this case, we set $\tau_R = 0.5T_2^*$, such that each Ramsey sequence duration is $\tilde\tau = 0.5T_2^* + \tau_o$. Here plot $g^\omega$ as a function of the overhead time $\tau_o$ and the  \santi{inter-pulse time $\tau$ of the corresponding DD sequence}. Note that, in both figures, we choose the amount of Ramsey sequences $N$, that an experiment features, to be $N = \lfloor T/\tilde\tau \rfloor$, the integer part of the quotient, meaning that it will increase with the total measurement time, and decrease as either $\tau_R$ or $\tau_o$ grows. The DD sequence duration is, in both cases, independent of the correlated Ramsey parameters, and sets the frequency of the target signal, which is chosen in both figures to be $\omega = \pi/\tau$, defined in terms of the DD sequence inter-pulse time $\tau$. In both figures the dashed line indicates the boundary $g^\omega$ =  1, and we consider the amplitude $\xi = 1/T_2^*$ to be the same in both protocols. For slow frequencies, a correlated Ramsey solution can always be found which \santi{in principle} outperforms DD sequences.
   }\label{Fig2}
\end{figure}

Eq.~(\ref{lowfreqgain}) shows that $g^\alpha$ grows exponentially for frequencies $\omega < \pi/T_2$ that require $T > T_2$, explicitly showing the physical connection between correlated Ramsey and dynamical decoupling, and explaining why the former performs well in the low frequency regime. As the DD sequence duration approaches the coherence time of the probe, it is not possible to disregard its influence in shortening the Bloch vector, reflected as an exponential decay of the estimation precision. Correlate Ramsey on the other hand, permits adjusting the sequence duration $\tau_R$ such that the Bloch vector is renewed in every measurement without a significant loss in its length, as it happens in high frequency dynamical decoupling, whose limit it recovers. In Fig.~\ref{Fig2} we use $T_2^*$ as unit of measure, and calculate the frequency gain $g^\omega$ {\textit{exactly}}, using Eq.~(\ref{eq:exactgain}) with the accumulated phases Eqs.~(\ref{acphaser1}) and (\ref{acphasehn}), assuming the frequency matching condition $\omega = \pi/\tau$ for DD. The approximate expression Eq.~(\ref{lowfreqgain}), which reproduces well the trends observed in the exact results, as shown in Fig.~\ref{2Dplot} in Appendix \ref{Methodscomparison}, can then be used to gain insight into the behaviour of each protocol in the low frequency regime.

The probe coherence time is usually modelled as a function that grows sublinearly with the number of pulses $M$ \cite{deLange2010}. In Fig.~\ref{Fig2}(a) we favour DD considering that $T_2 = MT_2^*$ (with $M=8$), and vary $\tau_R$ aiming for detection of a signal with $\omega < \pi/T_2$. 
We observe that it is always possible to find a $\tau_R$ such that correlated Ramsey improves the results from DD whenever $T > T_2$. For frequencies that can be fit within the coherence time $T_2$, but which still remain small enough (such that $T$ must be larger than the dephasing time $T_2^*$), in Fig.~\ref{Fig2}(b) we show that, for $\tau_R$ to $0.5T_2^*$, it  \santi{can} still \santi{be} advantageous to  choose to perform correlated Ramsey over the typical \santi{DD} choice, even in an optimistic coherence time scenario for DD in which $T_2 = 30T_2^*$, \santi{conditioned to being able to keep the overhead time short}. Thus, these results map the transition from a high frequency regime in which DD is preferred, to a low frequency regime, defined by the coherence time of the probe, in which correlated Ramsey has a better performance.

\subsubsection{Sequence optimisation}

The Fisher information is an objective measure of the performance that a measurement protocol will show on an experiment, as it defines the minimum error that will be observed on a parameter estimation derived from a given experimental data. The similarity between the exact numerical results and the approximate expression Eq.~(\ref{lowfreqgain}), reflects that the approximate formulas for the Fisher information accurately mirror the behaviour of each protocol according to the defined sequence parameters. Then, we can use Eq.~(\ref{FIRamsey}) to  derive the optimal correlated Ramsey sequence parameters for a given experiment. To do so, we rewrite Eq.~(\ref{FIRamsey}) in terms of the sequence parameters, replacing $T$ by $N(\tau_R + \tau_o)$. Since the total Fisher information is accumulative, we can drop the $N$, which is a global factor that will not affect the position of the maximum, and which only tells that, the more measurements performed, the better the estimation will be. Thus, we are left with
\be
I_{NR}^\alpha(\tau_R,\tau_o,T_2^*) = f_{NR}^\alpha[(\tau_R + \tau_o)]\frac{\tau_R^2}{2}e^{-2 \tau_R/T_2^*},
\label{optimiseI}
\ee
where $f_{NR}^\omega[(\tau_R + \tau_o)] = \xi^2(\tau_R + \tau_o)^2/3$,  $f_{NR}^\xi=1$ and  $f_{NR}^\varphi=\xi^2$.

Of the parameters susceptible of optimisation, the dephasing time $T_2^*$ is obvious, as the longer it is, the smaller the estimation error will be. However, it is relatively fixed by the probe that it is being used, and by the type of experiment considered. Something similar happens to the overhead time $\tau_o$, which, from Fig.~\ref{Fig2}(b), it is required to be as short as possible. Nonetheless, $\tau_o$ is lower-bound by the requirement of measuring the qubit and reinitialising it with sufficiently high fidelity, being thus limited by the experiment characteristics. Then, the only parameter that can be tailored to obtain the maximum information, is the phase acquisition time per Ramsey sequence $\tau_R$. From  Eq.~(\ref{optimiseI}) we see that said optimal $\tau_R$ depends on the parameter to be estimated. Then, for the amplitude $\xi$ or the phase $\varphi$, the optimal measurement strategy is to choose $\tau_R = T_2^*$
In the case of frequency estimation, there are two possible solutions for $\tau_R$ that yield a local maximum of the Fisher information: 
\be
\tau_R^{\pm} = T_2^* - \frac{\tau_o}{2} \pm \sqrt{\frac{\tau_o^2}{4} +
{T_2^*}^2}.
\label{optimaltaur}
\ee
Of these two options, $\tau_R{-}$ leads to a negative time which, if taken as an absolute value, does not yield a maximum. This leaves us with $\tau_R^+$ as the optimal phase acquisition time for correlated Ramsey. Surprisingly, we can see that, optimising the overhead time, which means making it as small as possible, yields an optimal $\tau_R \rightarrow 2T_2^*$. Conversely, for a large overhead time $\tau_o \gg T_2^*$, the optimal Ramsey sequence measurement time for frequency estimation is $\tau_R\rightarrow T_2^*$.

These results imply two important things: First, both the optimal overhead time, and, crucially, the optimal Ramsey measurement time, are substantially different from those chosen when the probe features long dephasing times, in excess of ms
\cite{Barry2020,Herbschleb2019,TheOther}, mainly due to both the overhead time and the dephasing time being similar, which means none of them can be disregarded during the optimisation procedure. And second, Eq.~(\ref{optimaltaur}) shows that, in fact, rather surprisingly, the best strategy for optimal frequency information acquisition is to enlarge each measurement beyond the dephasing time of the probe.

\subsection{Detection and sensitivity}

\begin{figure*}\includegraphics[width=13cm]{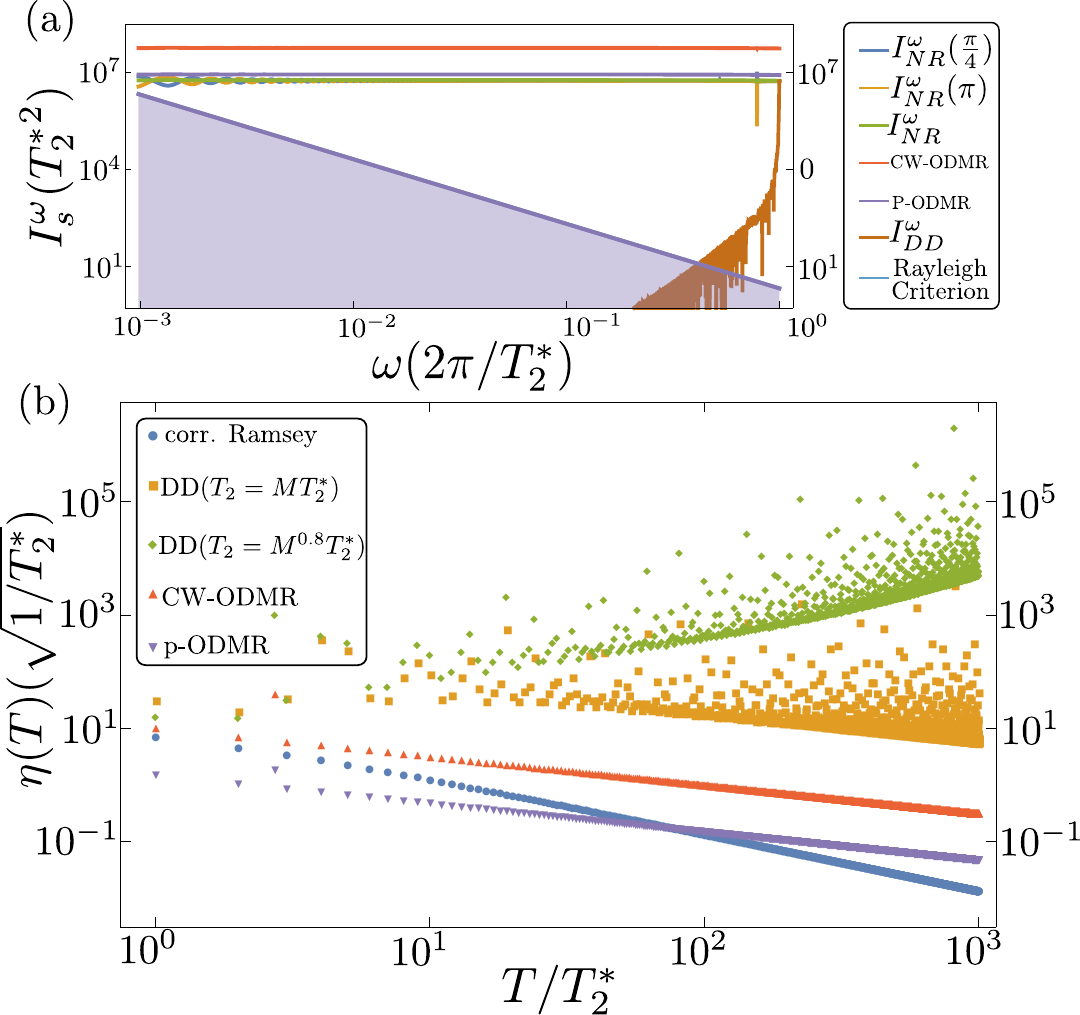}
    \caption{(a) Fisher information for frequency estimation $I_{NR}^\omega$, for a correlated Ramsey sequence of duration $T = 1000T_2^*$, for two choices of $\varphi$ and the phase-averaged result. $I_{DD}^\omega$ for a DD sequence where we assume that the pulses provide a linear coherence time enhancement $T_2 = MT_2^*$, and we define $M = \lceil T\pi/\omega \rceil$, the integer part of the quotient plus one, to ensure that, at least, $M=1$. Fisher information corresponding to both continuous wave (CW-ODMR) and pulsed (p-ODMR) optically detected magnetic resonance, adjusted from \cite{Dreau2011,Zhang2021,SEGAWA202320}.
    The shadowed area marks the region where $I_s^\omega$ is not sufficient for frequency estimation, set here at $\Delta\omega < \omega^2/4 $ as per the Rayleigh criterion, showing that only correlated Ramsey can successfully access the low frequency regime. (c) Frequency detection sensitivity $\sqrt{T/I_s^\omega}$ as a function of the total measurement time, calculated in the low frequency regime for a target signal of frequency $\omega = \pi/T_2^*$, for correlated Ramsey, a DD sequence with either $T_2 = MT_2^*$ or $T_2 = M^{0.8}T_2^*$, CW-ODMR and p-ODMR. Note that all $I_s$s are calculated exactly and that $\tau_R = \tau_o = 0.5T_2^*$ in all Figures, where we choose $\xi = 1/\tau_R$ ($\xi = 1/\tau$ for DD), which is optimal for parameter estimation \cite{Oviedo2020}. Smaller values of $\xi$ would just require larger $T$. In the case of ODMR we consider optimal detection efficiencies and contrast \cite{SEGAWA202320}.}\label{Fig3}
\end{figure*}

We now turn to the specific frequency detection capabilities of correlated Ramsey, and compare them to various state-of-the-art methods for low frequency sensing with quantum probes, namely, continuous wave optically detected magnetic resonance (CW-ODMR) and its pulsed sibling (p-ODMR) \cite{Acosta2010,Schoenfeld2011,Dreau2011,Clevenson2015,Schloss2018}. Fig.~\ref{Fig3}(a) we calculate the exact $I_{s}^\omega $ of frequency estimation for a total measurement time $T = 1000T_2^*$, and compare it with the minimal amount of information --lying above the shadowed region-- that is required for a successful parameter estimation, defined by the Rayleigh criterion (RC) of classical optics to be $I_s^\omega > 4/\omega^2 = 1/\Delta\omega_{RC}$ \cite{Abbe1873,Rayleigh1879,JonesAR1995}. We calculate $I_{NR}^\omega $ for two instances of the phase $\varphi$ of the signal, and averaging out such phase. While particular phases cause oscillations of $I_{NR}^\omega$ at low frequencies, an experiment which measures several time-traces with arbitrary phases on each of them \cite{Rotem2019} shall have no problem in estimating any arbitrarily low frequency, provided the measurement time $T$ is sufficiently long. As expected, $I_{DD}^\omega$ decreases dramatically for low frequencies. The comparison with the two ODMR variants is calculated exactly by adapting the results from Refs.~\cite{Dreau2011,Zhang2021,SEGAWA202320} to get a measure of the Fisher information from the mean squared error for optimal photon counting and contrast, and the same total measurement time $T$. We can see that while CW-ODMR is relatively limited by the low sensitivity inherent to the technique \cite{SEGAWA202320}, pulsed ODMR achieves similar results as correlated Ramsey. It is noteworthy that all the viable protocols are valid for all low frequencies considered, showing unperturbed Fisher information regardless of the frequency.

Finally, in Fig.~\ref{Fig3}(b) we calculate the sensitivity to the frequency as a function of the total measurement time $T$, which following \cite{Barry2020} can be defined as $\eta_s = \sqrt{T/I_s^\omega}$. We include two instances of $T_2$ behaviour with DD $\pi$ pulses, namely $T_2 = MT_2^*$ and $T_2 = M^{0.8}T_2^*$, to which correlated Ramsey is far superior, even for a coherence time linearly growing with the number of pulses, which is the only viable DD protocol for low frequency sensing. As well, we include the sensitivity for both continuous wave and pulsed ODMR. We can see that pulsed ODMR can be a better low frequency detection scheme whenever the experiment time is constrained by different considerations (e.g. a short sample coherence time), but the superior scaling with the total measurement time displayed by correlated Ramsey with respect to ODMR --that is, $1/\sqrt{T^3}$ vs $1/T_2^*\sqrt{T^2}$-- means that correlated Ramsey can always achieve better sensitivities.

\subsection{Real sensors performance}

We have already demonstrated that performing correlated Ramsey measurements is an alternative, promising pathway for low frequency signal detection, on par or better than state-of-the-art sequences. Now, we focus on the correlated Ramsey sequence itself and its possible implementation on different quantum sensors.

A crucial feature of the correlated Ramsey sequence is that it decouples the parameters describing the signal from those of the probe or the sequence. That is, it does not require that any pulses spacing matches the half period of the external signal, meaning that it becomes possible to optimize the sequence without having prior information about the target signal, based solely on knowledge from the sensor. 

Following this intuition, we can calculate the expected performance of various sensors with different signals. For example, for one of the most prominent nano-scale magnetometers, the NV center in diamond, following the results from the previous section, the sensitivity of correlated Ramsey to a magnetic signal is defined as 
\be
\eta_{NR} = \frac{\hbar}{\mu_B n_{\text{avg}} C}\sqrt{\frac{T}{I^\omega_{NR}}},
\label{CRsensitivity}
\ee
with $C$ the measurement contrast and $n_{\text{avg}}$ the average number of photons detected per measurement. Thus, assuming a $T_2^* $ = 2 $\mu$s, considering a sub-optimal $\tau_R = T_2^*/2$, and taking a fairly typical measurement contrast of 30\% and an average photon number $n_{\text{avg}}=150$ kcounts/s we have that, for an overhead time $\tau_o = T_2^*/2$, the sensitivity to a signal of frequency $\omega = \pi/T_2$ with $T_2$ = 100 $\mu$s is $\eta_{NR}$ = 2 $nT/\sqrt{\text{Hz}}$. Note that this sensitivity is independent of the frequency of the signal being probed as long as $\omega < \pi/T_2^*$, as can be deduced from Fig.~\ref{Fig2}(a), whereas for larger frequencies it quickly grows, becoming impractical for sensing. Furthermore, note that this sensitivity can be improved by choosing an optimal $\tau_R \approx$ 1.8 $\mu$s, as per Eq.~(\ref{optimaltaur}). As a comparison, for similar parameters, a pulsed ODMR experiment achieves a sensitivity $\eta_{p-ODMR}$ = 8 $nT/\sqrt{\text{Hz}}$ \cite{SEGAWA202320}.

Using Eq.~(\ref{CRsensitivity}) we can estimate the sensitivity that different quantum sensors can attain, with the correlated Ramsey sequence, for low frequency sensing. Alternative platforms such as trapped ions or Rydberg atoms feature much longer $T_2^*$ times (ms to s) than NV centers, at the prize of operating at very low or very high temperatures. The added advantage of an extended $T_2^*$ is the limited influence that the overhead time has on the sensor performance. Thus, for a trapped ion with 2 ms $T_2^*$ time, a sensitivity $\eta_{NR}$ = 4 $pT/\sqrt{\text{Hz}}$ for a signal of frequency 1 Hz. For the same frequency, but with a $T_2^*$ = 1 s on a Rydberg atom, the sensitivity can be decreased further to $\eta_{NR}$ = 10 $fT/\sqrt{\text{Hz}}$. These numbers put the correlated Ramsey sequence in line with state-of-the-art quantum sensing sequences performance, offering an operating regime that reaches frequencies inaccessible by conventional techniques for ac sensing.

\section{Discussion and Conclusions}
\label{sec:discussion}

We consider the correlated Ramsey protocol for quantum sensing of low frequency signals \cite{TheOther}. We demonstrate with exact numerical calculations that such a protocol behaves as a DD sequence attuned for the low frequency regime, avoiding the limitations imposed by the coherence time of the probe, \santi{thus outperforming the commonly used DD sequences in the regime where the coherence time of the probe is shorter than the half period of the signal}. We provide a simple but accurate analytical expression that permits estimating the adequate parameters to design tailored Ramsey sequences for any given target signal, show which are the optimal parameters for a given sequence, and demonstrate the achievable sensitivity for different quantum sensing platforms, comparing correlated Ramsey with \santi{the performance of} current state-of-the-art low frequency sensing schemes, \santi{which correlated Ramsey matches}. Importantly, information about the phase of the signal can be acquired by performing sequential measurements with a precise time separation. This is the key that permits Ramsey measurements to become sensitive to oscillating signals, even though it is originally intended for static signals. Signal processing for parameter estimation can be done through a variety of methods, including  maximum likelihood estimation \cite{Rotem2019}, or via least squares fitting to a signal's correlation model or Fourier transform \cite{Oviedo2020,Staudenmaier2022,Staudenmaier2023}.

The low frequency regime features prominently in a wide variety of fields, ranging from the study of J couplings in molecules, to the life sciences \cite{Blanchard2013,Clevenson2015,Barry2016,Arai2022}. Our analysis is made all the more compelling as we demonstrate that no frequency matching condition is required for correlated Ramsey, which also equals or improves the sensitivity with respect to other state-of-the-art low frequency sensing schemes based on  the optically detected magnetic resonance protocol. Correlated Ramsey reduces the number of pulses required per measurement, and limits the damage produced by laser-induced broadening. As such, the correlated Ramsey protocol is an essential addition to the low frequency sensing toolbox.

\emph{Acknowledgements} --- S.O.C. acknowledges the support from the María Zambrano fellowship, the Fundación Ramón Areces postdoctoral fellowship (XXXI
edition of grants for Postgraduate Studies in Life and Matter Sciences in Foreign Universities and Research centers) as well as the
Israeli Science Foundation and the ERC grant QRES, project number 770929. J.C. acknowledges support from Ministerio de Ciencia,
Innovación y Universidades (Spain) (‘‘Beatriz Galindo’’ Fellowship BEAGAL18/00078), from grant PID2021-124965NB-C22 funded by MICIU/AEI/10.13039/501100011033 and by "ERDF/EU", and from European Union project C-QuENS (Grant No. 101135359). J.P. acknowledges support from the QuantERA II Programme (Mf-QDS) that has received funding from the European Union’s Horizon 2020 research and innovation programme under Grant Agreement No 101017733, and from grant TED2021-130578B-I00 funded by MICIU/AEI/10.13039/501100011033 and by the "European Union NextGenerationEU/PRTR". J.P. also acknowledges funding received through the grant PID2021124965NB-C21 funded by MICIU/AEI/10.13039/501100011033.

\appendix
\onecolumngrid

\section{Phase accumulation}\label{Methodsphase}
We provide a detailed account on how to calculate the phase that a qubit superposition state accumulates when it is interacting with an external signal, and subject to either the Ramsey sequence or a generic dynamical decoupling sequence. We consider that both sequences feature an initial $\pi/2$ pulse that creates the superposition state on the qubit, $\Psi(0) = (\ket{0} + \ket{1})/\sqrt{2}$, and a final $\pi/2$ pulse after the evolution time which rotates the evolved state onto the measurement basis. Throughout, we consider that the pulses intended to control the qubit have negligible duration (impulsive limit), and that their amplitude is much smaller than the qubit energy gap.

\textit{Single Ramsey experiment} --- During the free evolution time $\tau_R$ in the Ramsey sequence, the quantum probe superposition state evolves influenced by the external signal $S(t,\omega,\xi,\varphi)$,  such that after some time $t$ the state of the probe is $\Psi(t) = \left(\ket{0} + \exp\left[i\Phi_R(t,\omega,\xi,\varphi)\right]\ket{1}\right)/\sqrt{2}$, with $\Phi_R(t,\omega,\xi,\varphi)$ a phase that the probe accumulates, and which upon transformation onto the measurement basis and interrogation of the qubit state, reveals itself as a population difference between the states $\ket{0}$ and $\ket{1}$. It is this population difference what carries the information about the external signal parameters. 

For our purposes here, we consider a pure tone signal of cosine form, such that $S(t,\omega,\xi,\varphi) = \xi\cos\left(\omega t + \varphi\right)$. Then, for a Ramsey measurement starting at $t=0$ and with duration $\tau_R$, the accumulated phase is calculated as  
\be 
\begin{split}
\Phi_R(\tau_R,\omega,\xi,\varphi) &= \xi \int_{0}^{\tau_R} dt\cos\left(\omega t + \varphi\right) \\& =  \xi\tau_R\cos\left(\frac{\omega\tau_R}{2} + \varphi\right){\text{sinc}}\left(\frac{\omega\tau_R}{2}\right).\label{RamseyphaseSI} 
\end{split}
\ee

The accumulated phase Eq.~(\ref{RamseyphaseSI}) can be rewritten as $\xi\Re\left\{e^{i\varphi}{F}\left[h_R(t)\right] \right\}$, the real part of the Fourier transform (or, alternatively, the cosine Fourier transform) of a function $h_R(t) = 1$ in $0 \leq t \leq \tau_R$ and zero otherwise. Such a function is called \textit{response function}, and can be said to characterize the value of the coherence of the qubit probe during the evolution time. Describing the accumulated phase in terms of the Fourier transform of response functions proves useful for more complicated measurement sequences, as will be shown in the next subsection for dynamical decoupling.

To demonstrate that it is sensitivity to the phase of the external signal $\varphi$ what allows Ramsey measurements to detect oscillating signals, we can solve Eq.~(\ref{RamseyphaseSI}) analytically to find its maximum in the frequency $\omega$. This yields \be
\omega\tau_R = \tan(\omega\tau_R + \varphi) - \frac{\sin\varphi}{\cos(\omega\tau_R + \varphi)}.
\label{maxRamsey}\ee For $\varphi \neq n\pi$ the maximum phase accumulation occurs at a frequency $\omega \neq 0$. 

The connection with usual experiments, in which the Ramsey sequence is mostly sensitive to static signals, can be made by squaring Eq.~(\ref{RamseyphaseSI}), and averaging it with respect to the phase to get the filter function, which is the typical figure of merit for signal detection. Then
\be
\begin{split}
\langle \Phi_R(\tau_R,\omega,\xi,\varphi)^2\rangle_{\varphi} &= \xi^2\tau_R^2{\text{sinc}}^2\left(\frac{\omega\tau_R}{2}\right)\frac{1}{2\pi}\int_{0}^{2\pi}d\varphi\cos^2\left(\frac{\omega\tau_R}{2} + \varphi\right) \\& = \frac{\xi^2\tau_R^2}{2}{\text{sinc}}^2\left(\frac{\omega\tau_R}{2}\right),\label{phaseaverageSI}
\end{split}
\ee
whose maximum occurs at $\omega = 0$.

\textit{Dynamical decoupling} --- Dynamical decoupling sequences are based on the Hahn echo of classical nuclear magnetic resonance \cite{Cywinski2007}. As such, they comprise a number $M$ of $\pi$ pulses acting on the qubit probe during the evolution time, i.e. in between the two $\pi/2$ pulses that initiate and end the measurement sequence. These $\pi$ pulses are usually evenly spaced in time, with time separation $\tau$, and their intention is to rephase the qubit state. Specifically, when acting on the state $\Psi(t,\omega,\xi,\varphi)$, they will invert the state with respect to a particular axis of the Bloch sphere. 

To calculate the phase accumulated during a dynamical decoupling sequence, we again have to integrate the external signal for the duration of the evolution time, which for $M$ $\pi$ pulses applied is $M\tau$, where we consider that the separation between the first $\pi$ pulse and the initial $\pi/2$ pulse is $\tau/2$ (see Fig.~1 on the main text for a depiction of both protocols). The same happens for the last $\pi$ pulse and the $\pi/2$ that finishes the sequence. Moreover, we have to consider the specific effect that each of the $\pi$ pulses has on the qubit. To do so, we resort to the response function, and start off from the Ramsey sequence case, which is also valid to describe the value of the qubit coherence during a free evolution time $\tau$. Then, since $\pi$ pulses reverse the sign of the qubit coherence, a dynamical decoupling sequence can be constructed using response functions of absolute value $1$ and that change sign every time a $\pi$ pulse is applied. For example, the case of a Hahn echo of duration $\tau$, with a $\pi$ pulse applied at a time $\tau/2$ is described as $h_1(t) = h_R(t) - h_R(t - \tau/2)$, where we have taken $\tau_R = \tau/2$. The generic dynamical decoupling sequence response function is constructed repeating the Hahn echo sequence every $\tau$, such that $h_M(t) = \sum\limits_{j=0}^{M-1}(-1)^jh_1(t-j\tau)$. It is in this situation where writing the accumulated phase as a Fourier transform proves useful, as we can then apply the time shift property of the Fourier transform, that reads $\mathcal{F}[h(t-t')] = \exp{(i\omega t')}\mathcal{F}[h(t)]$, to the Ramsey sequence results, and calculate the accumulated phase on a generic dynamical decoupling sequence \cite{Cerrillo2021}. Beginning with the Hahn echo sequence and generalizing we get
\be
\begin{split}
\Phi_{DD}(T,\omega,\xi,\varphi) &= -\xi\tau\cos\left(M\frac{\omega\tau+\pi}{2} + \varphi \right) \\&
\frac{\sin\left(M\frac{\omega\tau+\pi}{2}\right)}{\cos\left(\frac{\omega\tau}{2}\right)}\sin\left(\frac{\omega\tau}{4}\right){\text{sinc}}\left(\frac{\omega\tau}{4}\right).
\label{DDphaseSI}
\end{split}
\ee

\section{Small detuning $\delta$ limit for dynamical decoupling}

We now calculate explicitly the small detuning limit $\delta \rightarrow 0$ of the accumulated phase $\Phi_{DD}(T,\omega,\xi,\varphi)$ in a dynamical decoupling sequence with $\omega_{DD} \approx \omega$. Consider the full expression for $\Phi_{DD}(T,\omega,\xi,\varphi)$ in Eq.~(\ref{DDphaseSI}). If the frequency matching condition is met, such that $\tau$ is chosen to satisfy $\pi/\tau = \omega_{DD} \approx \omega$, as it is generally required for a successful signal detection with dynamical decoupling sequences, we can write $\omega_{DD} = \omega - \delta$, with $\delta \ll \omega$ a small detuning frequency. Then, using $\omega_{DD} = \pi/\tau$, the following holds: $\omega\tau = \omega\pi/(\omega - \delta) = \pi/(1-\delta/\omega) \approx \pi + \pi\delta/\omega + O(\delta^2) \approx \pi + \delta\tau$, where in the last two steps we have made use of the small detuning assumption. A similar reasoning shows that 
$M\omega\tau \approx M(\pi + \delta\tau)$. Considering the terms which do not depend on $M$ in Eq.~(\ref{DDphaseSI}), we can write them as
\be
\begin{split}
&\frac{4}{\omega\tau}\frac{\sin^2\frac{\omega\tau}{4}}{\cos\frac{\omega\tau}{2}} = \frac{2}{\omega\tau}\frac{1-\cos\frac{\omega\tau}{2}}{\cos\frac{\omega\tau}{2}} \approx \frac{2}{\omega\tau}\frac{1 - \cos\frac{\pi}{2\left(1 - \frac{\delta}{\omega}\right)}}{\cos\frac{\pi}{2\left(1 - \frac{\delta}{\omega}\right)}} \\& \approx \frac{2}{\omega\tau}\frac{1 + \frac{\pi\delta}{2\omega} + O(\delta^2)}{-\frac{\pi\delta}{2\omega} + O(\delta^2)} \approx -\frac{4}{\pi\delta\tau},
\end{split}
\ee
where the last steps involve a Taylor expansion around $\delta = 0$, keeping only the leading order in $\delta$. The terms that contain $M$ are rewritten as
\begin{equation}
\begin{split}
& -\cos\left(M\frac{\omega\tau+\pi}{2} + \varphi \right) \sin\left(M\frac{\omega\tau+\pi}{2}\right) \\& \approx -\cos\left(\frac{M\delta\tau}{2} + \varphi \right)  \sin\left(\frac{M\delta\tau}{2}\right).
\end{split}
\end{equation}
Bringing all together with a bit of trigonometry gives us that the following expression for the accumulated phase as a function of the detuning $\delta$:
\begin{equation}
\Phi_{DD}^\delta(T,\omega,\xi,\varphi) = \frac{2M\xi\tau}{\pi}\cos\left(\frac{M\delta\tau}{2} + \varphi\right) \\ {\text{sinc}}\left(\frac{M\delta\tau}{2}\right),\label{SIeqdelta}
\end{equation}
which coincides (up to a $2/\pi$ factor) with Eq.~(\ref{RamseyphaseSI}) by replacing the sequence duration $M\tau \rightarrow \tau_R$ and $\delta \rightarrow \omega$.

\section{Detailed Fisher information calculations}\label{MethodsFI}

Here we provide a step by step derivation of the total Fisher information formulas, for a correlated Ramsey protocol and a dynamical decoupling sequence of equivalent total duration, in the low frequency limit, with which the approximate gain is calculated on the main text. We start from the expression for the Fisher information for a single instance of a general measurement sequence $s$
\be
I_{s}^\alpha = \frac{\sin^2\left[ \Phi_S(t,\omega,\xi,\varphi)  \right]}{\exp\left(2 t/T_s\right) - \cos^2\left[ \Phi_S(t,\omega,\xi,\varphi) \right]}\left[ \frac{d\Phi_S(t,\omega,\xi,\varphi)}{d\alpha} \right]^2,
\label{FIphaseSI}
\ee
with $\Phi_S(t,\omega,\xi,\varphi)$ the accumulated phase during that sequence and $\alpha$ any of the signal parameters. Focusing on the first factor, we can see that it oscillates between zero and $\exp\left(-2 t/T_s\right)$. We want to compare the information accumulated due to Ramsey measurements with that accumulated by a dynamical decoupling sequence. For long measurement time $T$ as required by dynamical decoupling sequences intended for low frequency signals, we can upper bound this first factor by $\exp\left(-2 T/T_2\right)$, and we do the same for correlated Ramsey sequences, thus, the following approximate expressions for $I_s^\alpha$ represent upper bounds on the Fisher information.
Additionally, we note that, in the low frequency limit, with $\omega \rightarrow 0$, ${\text{sinc}}(\omega\tau/2) \rightarrow 1$, while for the same reason the derivative ${\text{sinc}}'(\omega\tau/2) \rightarrow 0$, therefore, we can safely neglect the terms including ${\text{sinc}}'(\omega\tau/2)$ in the calculation. This works for a Ramsey measurement. In the case of a dynamical decoupling sequence, the argument is the same but for the approximate formula Eq.~(\ref{SIeqdelta}), in which, when the frequency matching condition of dynamical decoupling sequences is met, $\delta \rightarrow 0$. 

We begin by calculating the Fisher information of correlated Ramsey. For an arbitrary Ramsey sequence we have that 
\be
I_R^\alpha(\tau_R) = \exp(-2\tau_R/T_2^*)\left[ \frac{d\Phi_R(\tau_R,\omega,\xi,\varphi)}{d\alpha} \right]^2.
\ee
The total Fisher information in a sequence of $N$ measurements is obtained by summing the individual contributions from each Ramsey measurement composing the sequence. Notice that on implementing the correlated Ramsey protocol, we have to consider that at the end of each measurement the qubit is repolarized into its $\ket{0}$ state, and consequently the dephasing affects each measurement individually. This means that on performing the sum of Fisher informations we can take the exponential term out of the summation, such that 
\be
I_{NR}^\alpha(T) = \exp(-2\tau_R/T_2^*)\sum_{j=1}^N \left[ \frac{d\Phi_R(t_j,\omega,\xi,\varphi)}{d\alpha} \right]^2.
\label{generalsumSI}
\ee
Here, $t_j$ defines the starting times of each Ramsey sequence in the protocol, with $j$ running from $1$ to $N$, and where the separation between consecutive measurements is such that $t_{j+1} = t_j + \tilde\tau$, with $\tilde\tau = \tau_R + \tau_o$ including the overhead time $\tau_o$ spent on interrogating the qubit probe at the end of each sequence, and reinitializing it for the next measurement. Considering a $t_j$ starting time modifies Eq.~(\ref{RamseyphaseSI}) adding an extra $\omega t_j$ to the initial phase $\varphi$.

We begin with the case of the amplitude $\xi$, and we consider $t_1 = 0$ such that $t_j = j\tilde\tau$. Then, using $N=T/\tilde\tau$, we get that:
\be
\begin{split}
\sum_{j=1}^N \left[ \frac{d\Phi_{R}(\tau_R,\omega,\xi,\varphi)}{d\xi} \right]^2 &= \tau_R^2 {\text{sinc}}^2\left(\frac{\omega\tau_R}{2}\right)\sum_{j=1}^{N} \cos^2\left(\omega t_j + \frac{\omega\tau_R}{2} + \varphi \right) \\& \approx \frac{\tau_R^2T\cos^2\varphi}{\tilde\tau},
\end{split}
\ee
where we have applied the low frequency limit $\omega \rightarrow 0$.

The case of the phase $\varphi$ is similar, yielding
\be
\begin{split}
\sum_{j=1}^N \left[ \frac{d\Phi_{R}(\tau_R,\omega,\xi,\varphi)}{d\varphi} \right]^2 &= \xi^2\tau_R^2 {\text{sinc}}^2\left(\frac{\omega\tau_R}{2}\right)\sum_{j=1}^{N} \sin^2\left(\omega t_j + \frac{\omega\tau_R}{2} + \varphi \right) \\& \approx \frac{\xi^2\tau_R^2T\sin^2\varphi}{\tilde\tau}.
\end{split}
\ee

The frequency $\omega$ contains a few more terms, but can be calculated in the same way:
\begin{align}
&\sum_{j=1}^N \left[ \frac{d\Phi_{R}(\tau_R,\omega,\xi,\varphi)}{d\omega} \right]^2 \\ &= \tau_R^2\xi^2 {\text{sinc}}^2\left(\frac{\omega\tau_R}{2}\right)\sum_{j=1}^{N} (t_j + \tau_R/2)^2\sin^2\left(\omega t_j + \frac{\omega\tau_R}{2} + \varphi \right)\\
&=\tau_R^2\xi^2{\text{sinc}}^2\left(\frac{\omega\tau_R}{2}\right) \sum_{j=1}^{N} \left( t_j^2 + \frac{\tau_R^2}{4} + t_j\tau_R \right) \sin^2\left(\omega t_j + \frac{\omega\tau_R}{2} + \varphi \right) \\
& \approx\tau_R^2\xi^2\sin^2\varphi\left[ \frac{\tilde\tau^2}{6}\left(N + 3N^2 + 2N^3 \right) + \frac{\tau_R^2 N}{4} + \frac{\tau_R\tilde\tau}{2}\left(N+N^2 \right) \right]\\
& \approx \tau_R^2\xi^2\sin^2\varphi\left[ \frac{\tilde\tau^2N^3}{3} + O(N^2) \dots \right] \approx \frac{\xi^2\tau_R^2T^3\sin^2\varphi}{3\tilde\tau},\label{truncate}
\end{align}
where we keep just the leading order in $N = T/\tilde\tau$.

We can further simplify the expressions above for the total Fisher information of each parameter by taking a phase average, as we did in Eq.~(\ref{phaseaverageSI}), which reduces each of them by a factor of 2. Then, we can write them in compact form as 
\be
I_{NR}^\alpha(T) = f_{NR}^\alpha\frac{\tau_R^2T}{2\tilde\tau}\exp(-2 \tau_R/T_2^*) ,
\ee
which corresponds to the expression provided on the main text, with $f_{NR}^\omega = \xi^2 T^2/3$, $f_{NR}^\xi = 1$ and $f_{NR}^\varphi = \xi^2$. 

The Fisher information for a dynamical decoupling sequence can be immediately calculated using Eq.~(\ref{SIeqdelta}). Note that in a dynamical decoupling sequence, the phase $\varphi$ of the external signal will be random, and the sequence will gather no information about it. This means that in order to fairly compare the information about the frequency and the amplitude gathered on a dynamical decoupling sequence, to that acquired for the same parameters on a correlated Ramsey protocol of equal duration, we need to consider the uncertainty about the phase by again taking the average with respect to it. Then, for a dynamical decoupling sequence of total duration T, the Fisher information is
\be
I_{DD}^\omega(T) = \frac{\xi^2T^4}{2\pi^2} {\text{sinc}}^2\left(\frac{\delta T}{2}\right)e^{-2T/T_2} \approx f_{DD}^\omega\frac{T^2}{2\pi^2} e^{-2T/T_2},\label{IomegaDDSI}
\ee
for the frequency and 
\be
I_{DD}^\xi(T) = \frac{2T^2}{\pi^2} {\text{sinc}}^2\left(\frac{\delta T}{2}\right)e^{-2T/T_2} \approx f_{DD}^\xi\frac{T^2}{2\pi^2} e^{-2T/T_2},\label{IkappaDDSI}
\ee
for the amplitude. Then, $f_{DD}^\omega = 3f_{NR}^\omega$  while  $f_{DD}^\xi = 4f_{NR}^\xi$. Note that in both equations we consider the low detuning limit $\delta \rightarrow 0$.

\section{Short Hahn echo vs correlated Ramsey}\label{RamseyvsHahn}

\begin{figure}[!h]\centering\includegraphics[width=8cm]{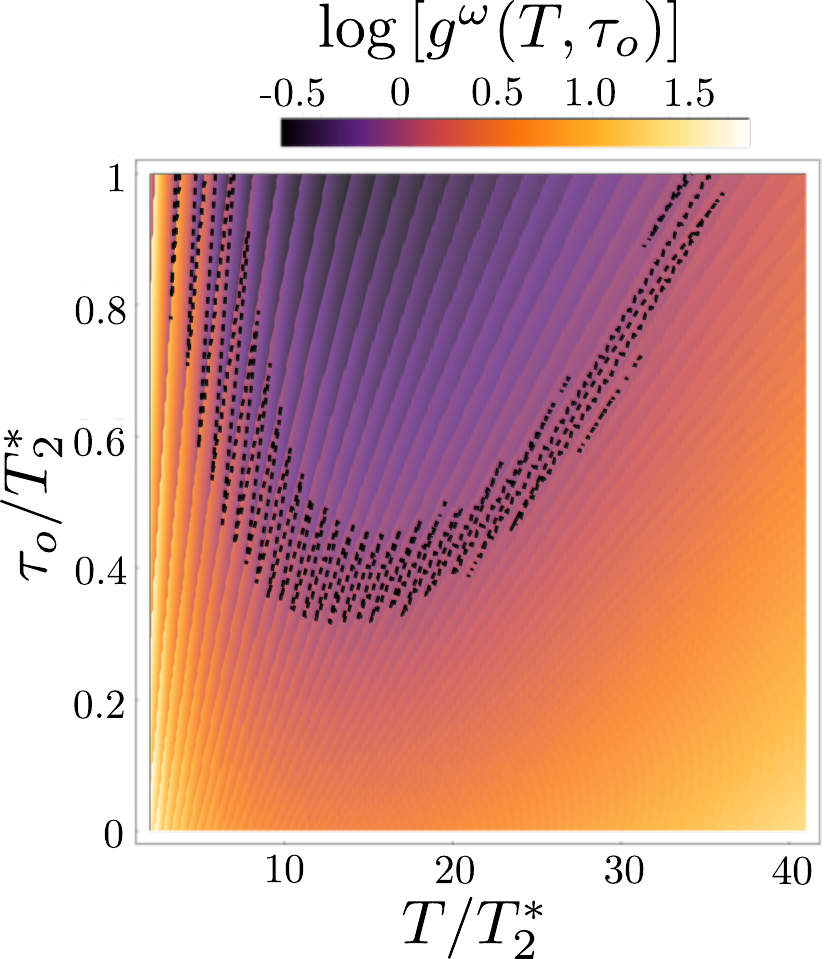}
    \caption{Logarithm of the exact gain for frequency estimation $g_\omega$ in Eq.~(\ref{eq:exactgain}). In this Figure, we compare correlated Ramsey to a Hahn echo sequence for which we assume a hypothetical strong recoherence effect of $T_2 = 30T_2^*$. The total length of the sequence $T = 2\tau$ ranges from $2T_2^*$ to $40T_2^*$, and the pulses separation $\tau$ for the Hahn echo sets the frequency of the target signal, which is chosen to be $\omega = \pi/\tau$. Additionally, we set $\tau_R = 0.5T_2^*$, such that each Ramsey sequence duration is $\tilde\tau = 0.5T_2^* + \tau_o$, and plot $g^\omega$ as a function of the overhead time $\tau_o$ and the duration of either experiment $T$. Note that we choose the amount of Ramsey sequences $N$, that an experiment features, to be $N = \lfloor T/\tilde\tau \rfloor$, the integer part of the quotient, meaning that it will increase with the total measurement time, and decrease as $\tau_o$ grows. The Hahn echo sequence duration is independent of the correlated Ramsey parameters, the dashed line indicates the boundary $g^\omega$ =  1, and we consider the amplitude $\xi = 1/T_2^*$ to be the same in both protocols. For slow frequencies, a correlated Ramsey solution can always be found which \santi{in principle} outperforms any Hahn echo.
    }\label{Fig2SI}
\end{figure}

In this Section, we compare the performance of correlated Ramsey with that of a Hahn echo. As we did for Fig.~\ref{Fig2}(b) in the main text, here we also choose an hypothetical probe for which the coherence time in the presence of $\pi$ pulses is $T_2 = 30T_2^*$, unaffected by the number of pulses, but in Fig.~\ref{Fig2SI}, rather than optimizing the DD sequence for said $T_2$ adjusting the number of pulses, we keep the Hahn echo sequence, which defines the total duration of the measurement to be $T= 2\tau$, i.e. twice the duration of the sequence. As well, the target signal frequency is defined through the pulses separation as $\omega = \pi/\tau$. Thus we explore the regime in which frequencies can still be fit within the coherence time $T_2$, but which remains small enough (such that $T$ must be larger than the dephasing time $T_2^*$). In this intermediate regime, despite the Hahn echo phase accumulation being relatively untouched by decoherence, especially at short measurement times, we see that choosing $\tau_R$ to be $0.5T_2^*$, it is still advantageous to perform correlated Ramsey, whenever the overhead time can be also kept short. This somewhat counter-intuitive behaviour is explained by the fact that the DD sequence of choice does not use the whole of the coherence time available efficiently, and due to the correlated Ramsey sequence being specifically designed for frequency detection, where accumulating measurements and correlating in post-processing increases the information that can be extracted. Thus, these results map the transition from a high frequency regime in which DD is preferred, to a low frequency regime, defined by the coherence time of the probe, in which correlated Ramsey has a better performance.

\section{Approximate gain comparison}\label{Methodscomparison}

\begin{figure}[!h]\includegraphics[width=\columnwidth]{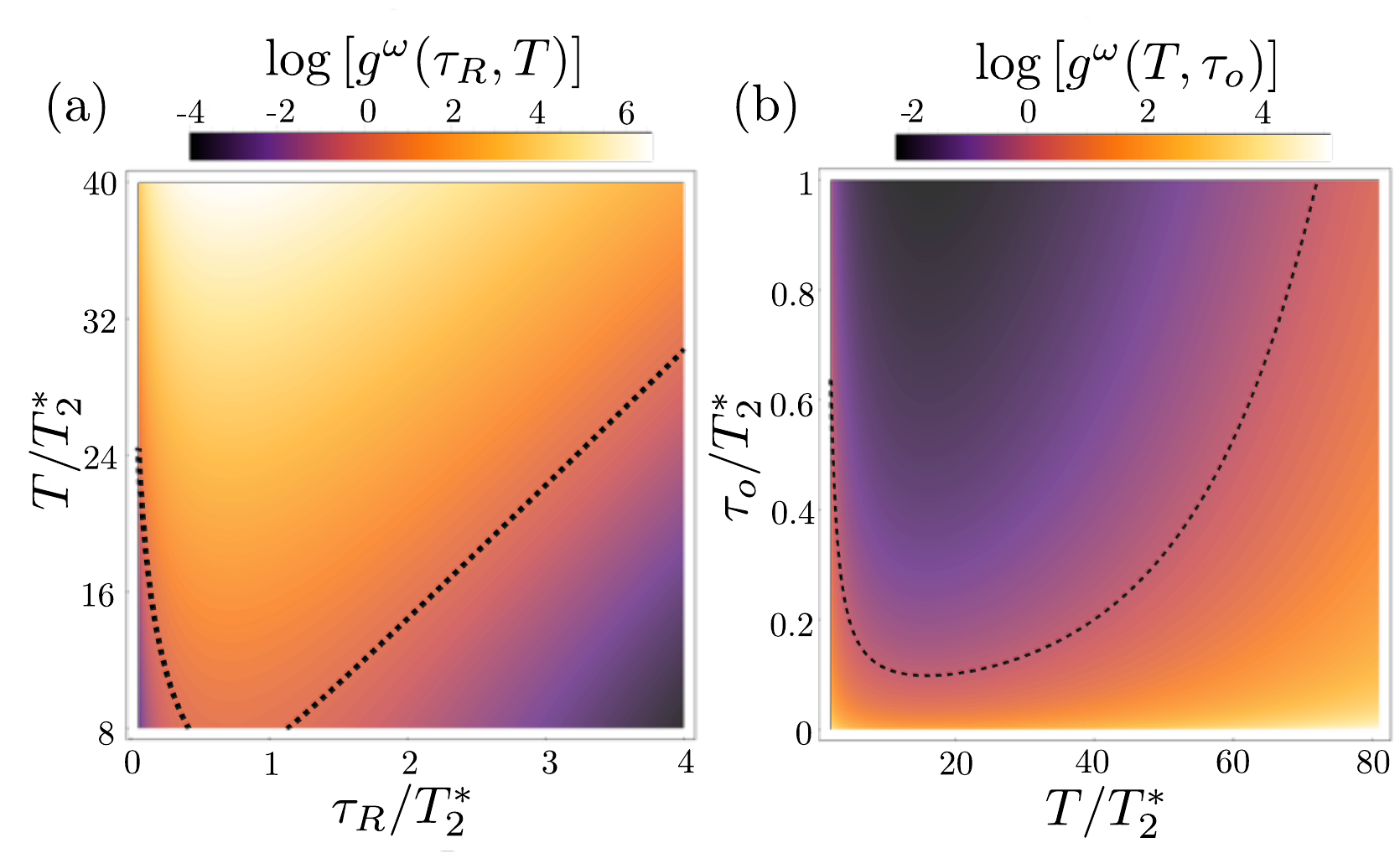}
    \caption{Logarithm of the approximate gain for frequency estimation $g_\omega$ in Eq.~(\ref{gainSI}). In (a) we compare correlated Ramsey to a DD sequence of eight pulses, which we assume to provide a coherence time enhancement $T_2 = 8T_2^*$. We set the duration of a single Ramsey measurement to $\tilde\tau = 0.5T_2^* + \tau_R$, which includes both the phase acquisition time $\tau_R$ and the overhead time $\tau_o$. We plot $g^\omega$, the gain corresponding to the frequency, as a function of $T$, the duration of either experiment, and of $\tau_R$. (b) Comparison between correlated Ramsey and a Hahn echo sequence for which we assume a hypothetical strong recoherence effect of $T_2 = 30T_2^*$. The total length of the sequence $T = 2\tau$ then ranges from $2T_2^*$ to $40T_2^*$. In this case, we set $\tau_R = 0.5T_2^*$, such that each Ramsey sequence duration is $\tilde\tau = 0.5T_2^* + \tau_o$. In this case, we plot $g^\omega$ as a function of the overhead time $\tau_o$ and the duration of either experiment $T$. Note that in both figures the dashed line shows the boundary $g^\omega$ = 1, and we consider the amplitude $\xi$ to be the same in both protocols, such that $g^\omega$ does not depend on $\xi$.
    }\label{2Dplot}
\end{figure}

We show now that the results for the gain presented on Fig.~\ref{Fig2}, which were calculated exactly, are well reproduced with the approximate formula 
\begin{equation}
    \lim_{\omega\rightarrow0}g^\alpha=e^{2\left(\frac{T}{T_2}-\frac{\tau_R}{T_2^*}\right)}\frac{\pi^2\tau_R^2}{T\tilde\tau}f^\alpha,\label{gainSI}
\end{equation}
with $f^\omega=1/3$ and $f^\omega=1/4$. We can see that the trends observed in the exact results are well captured by the approximate formula Eq.~(\ref{gainSI}), despite deviations [especially in Fig.~\ref{2Dplot}(b)], that can be attributed to the different approximations considered, specially truncating the number of terms with $T$ on Eq.~(\ref{truncate}), which, as all other approximations taken, favours dynamical decoupling sequences, in spite of which most scenarios of low frequency still show the advantage of performing correlated Ramsey measurements. 

The results presented here show that, despite the approximations taken, in the low frequency limit, the approximate formula Eq.~(\ref{gainSI}) can be used to estimate in a simple way the performance of a given correlated Ramsey protocol as compared to the equivalent dynamical decoupling sequence, allowing to choose the best Ramsey sequence parameters that maximize the gain.

\end{document}